\documentclass{article}
\usepackage[utf8]{inputenc}
\usepackage{amsmath}
\usepackage{amsfonts}
\usepackage{amssymb}
\usepackage{graphicx}
\usepackage{geometry}
\usepackage{hyperref} 
\hypersetup{
    colorlinks=true,
    linkcolor=blue,
    filecolor=magenta,      
    urlcolor=cyan,
    pdftitle={The Simulated Phoneme Speech Test},
    pdfpagemode=FullScreen,
}

\title{The Simulated Phoneme Speech Test (SimPhon Speech Test): A Data-Driven Method for \textit{In Silico} Design and Validation of a Phonetically Balanced Speech Test}
\author{Stefan Bleeck \\ Institute of Sound and Vibration Research (ISVR), University of Southampton, UK}
\date{13 June 2025}

\begin{document}

\maketitle

\begin{abstract}
Traditional audiometry often provides an incomplete characterization of the functional impact of hearing loss on speech understanding, particularly for supra-threshold deficits common in presbycusis. This motivates the development of more diagnostically specific speech perception tests. We introduce the Simulated Phoneme Speech Test (SimPhon Speech Test) methodology, a novel, multi-stage computational pipeline for the \textit{in silico} design and validation of a phonetically balanced minimal-pair speech test. This methodology leverages a modern Automatic Speech Recognition (ASR) system as a proxy for a human listener to simulate the perceptual effects of sensorineural hearing loss. By processing speech stimuli under controlled acoustic degradation, we first identify the most common phoneme confusion patterns. These patterns then guide the data-driven curation of a large set of candidate word pairs derived from a comprehensive linguistic corpus. Subsequent phases involving simulated diagnostic testing, expert human curation, and a final, targeted sensitivity analysis systematically reduce the candidates to a final, optimized set of 25 pairs (the SimPhon Speech Test-25). A key finding is that the diagnostic performance of the SimPhon Speech Test-25 test items shows no significant correlation with predictions from the standard Speech Intelligibility Index (SII), suggesting the SimPhon Speech Test captures perceptual deficits beyond simple audibility. This computationally optimized test set offers a significant increase in efficiency for audiological test development, ready for initial human trials.
\end{abstract}

\section{Introduction}
Traditional audiometry, while fundamental for quantifying hearing loss, often provides an incomplete picture of an individual's ability to understand speech, especially in challenging environments (Kricos, 2006; Bleeck, 2025). The core idea and initial analysis for an ASR-based frequency-specific speech test for diagnosing presbycusis are derived from Bleeck (2025); however, this work expands significantly by analyzing \textit{all} words in the English language, moving beyond fixed sets of random words. A significant limitation is the poor correlation between pure-tone audiogram (PTA) results and self-reported hearing difficulties or performance on speech perception tests. This discrepancy is particularly evident in supra-threshold deficits, which involve distortions in sound processing above the hearing threshold, such as reduced frequency resolution, impaired temporal processing, and altered loudness perception (Oxenham, 2008; Plomp, 1978; Wilson \& McArdle, 2005). These issues are not captured by simple audibility measures.

It is crucial to recognise that for mild-to-moderate hearing loss, a substantial part of the experienced difficulty, particularly in quiet or low-noise conditions, can often be attributed primarily to audibility. This refers to the simple inability to detect certain speech sounds, especially low-intensity, high-frequency consonants (e.g., /s/, /f/, /t/) which are critical for speech clarity (Kricos, 2006; Oxenham, 2008). This 'attenuation' aspect, as described by Plomp (1978), frequently represents the initial barrier to communication. However, supra-threshold distortions such as reduced frequency selectivity, impaired temporal processing, and altered neural coding are also inherent consequences of sensorineural hearing loss and play a significant role (Plomp, 1978; Wilson \& McArdle, 2005). While audibility may be the dominant factor in quiet and for mild losses, the impact of supra-threshold deficits becomes increasingly critical as background noise increases and as hearing loss progresses into the moderate range. In these challenging conditions, even when speech is amplified to be technically audible, these distortions can become the primary limiting factor preventing clear speech understanding (Wilson \& McArdle, 2005). Nonetheless, a strong, predictable link exists between specific patterns of frequency-specific audibility loss and the resulting phonetic confusions (Jean et al., 2025). This provides a compelling rationale for developing tests that can map these audibility-driven confusions, complementing traditional audiometry by offering valuable insights into an individual's specific perceptual challenges.

Recent advancements in Automatic Speech Recognition (ASR) systems, particularly those built on deep neural networks, offer a promising avenue to overcome these limitations. ASR can automate speech audiometry with accuracy and reliability comparable to human manual scoring, significantly reducing clinician time (Hnath \& Van Handel, 2016; Levenshtein, 1966). Crucially, ASR facilitates phoneme-level scoring, providing more detailed data on perception errors by identifying specific phoneme confusions rather than just whole-word misses (Humes, 2007). This capability allows for a more fine-grained, frequency-specific diagnostic insight into speech perception deficits. This paper details the development of a novel ASR-based frequency-specific speech test designed to provide such insights, particularly for conditions like presbycusis. Our approach leverages ASR to simulate human listeners under controlled acoustic degradation, mimicking typical hearing loss conditions. By meticulously simulating typical hearing loss conditions and analyzing the resulting ASR confusion patterns at the phonetic level, we aim to create a test capable of highlighting specific frequency regions impacted by perceptual degradation, offering a "confusion profile" that complements traditional audiometric data.

\section{Methodology}
The Simulated Phoneme Speech Test (SimPhon Speech Test) was developed through a six-phase computational pipeline, meticulously implemented in MATLAB. This iterative process systematically reduced a vast number of potential test items to a final, curated set of 25 pairs. The initial pool of words, derived from the Carnegie Mellon University (CMU) Pronouncing Dictionary, comprised over 114,000 items after initial filtering for English alphabet characters. The core of the methodology is a "simulated listener" comprised of a hearing loss model and the \texttt{wav2vec2.0} Automatic Speech Recognition (ASR) model for speech transcription.

\subsection{Phase 1: Large-Scale Data Generation \& Phonetic Analysis}
The objective of the initial phase was to create a large, ecologically plausible dataset of word confusions that result from high-frequency hearing loss. Audio for words from the Carnegie Mellon University (CMU) Pronouncing Dictionary was synthesized using the native macOS \texttt{say} Text-to-Speech (TTS) engine. Each audio file was then processed through our simulated listener pipeline. A confusion was logged whenever the ASR output for the hearing-impaired simulation differed from the original input word. This process yielded a dataset of approximately 8,800 ASR-generated word confusions. A detailed phonetic analysis was subsequently performed on this dataset to identify the most frequent phoneme substitutions (e.g., S $\rightarrow$ F, T $\rightarrow$ D), which formed the empirical basis for the next phase.

\subsection{Phase 2: Data-Driven Candidate Item Curation}
Using the primary phonetic substitution patterns identified in Phase 1, a MATLAB script systematically searched the CMU dictionary for all real-word minimal pairs that exemplified these specific, high-frequency confusions. This data-driven approach ensured that the candidate items were directly linked to the error patterns produced by the simulated hearing loss. The process generated a large candidate list of over 55,000 potential word pairs for testing.

\subsection{Phase 3: \textit{In Silico} Validation \& Methodological Pivot}
The 55,000 candidate pairs were run through a validation pipeline to assess their basic diagnostic potential. A critical methodological pivot occurred here: an initial approach using a Python-based TTS proved unreliable, leading to the adoption of the native macOS \texttt{say} TTS command for its superior speed and stability in the MATLAB environment. A pair was considered "validated" if the simulated normal-hearing listener correctly identified the original word, while the simulated hearing-impaired listener identified the target confused word. This initial filtering process successfully validated a list of approximately 140 robust word pairs.

\subsection{Phase 4: Expert Curation (Human-in-the-Loop)}
While computationally validated, not all pairs are suitable for a clinical test due to factors like word obscurity, ambiguity, or awkwardness. A human-in-the-loop curation step was therefore essential. The project's principal investigator listened to all $\sim$140 pairs. The audio was presented under challenging conditions because the PI has normal hearing (a simulated moderate hearing loss at 0 dB SNR) to mimic the target listening scenario. The investigator judged each pair on its subjective quality and suitability for a clinical test, yielding a high-quality list of 58 "useful" pairs. Reasons to eliminate pairs were acoustic: some pairs were too obviously different because for example of the length of the stimulus, or the words were abstruse and not often used (`clek` vs `gleek`, `machala` vs `matala` ) not appropriate for a child friendly speech test (it is amazing how many rude word the English language has!).

\subsection{Phase 5: Final Diagnostic Simulation}
The 58 curated pairs were subjected to a final, rigorous diagnostic simulation to quantify their performance under harsh conditions. For this phase, a "mild" hearing loss profile was used (\[0, 0, 0, 20, 25, 30, 40\] dB attenuation) at 10 dB SNR. For each word in every pair, 50 trials were run for both a normal-hearing and hearing-impaired simulation.

To better simulate a 2AFC task, a "closest guess" algorithm using phonetic Levenshtein distance was employed (Levenshtein, 1966). The ASR's output on each trial was scored based on whether its phonemic string was closer to the original stimulus or the target confusion. This process generated a detailed trial-by-trial log of raw ASR outputs, which was subsequently used for the in-depth phoneme-level error analysis (deletions, insertions, and substitutions) as described in Section 3.2. This phase produced robust Sensitivity (true positive rate for the HI model) and Specificity (true negative rate for the NH model) scores for all 116 possible confusion directions, which were combined into a single Youden's J statistic (J=Sensitivity+Specificity-1) to rank each item by its overall diagnostic power.

\subsection{Phase 6: Phonetically Balanced Test Set Selection (SimPhon Speech Test-25)}
The final step was to select the SimPhon Speech Test-25 test set from the 116 ranked items using a diversity-enforced selection algorithm. This script selected the items with the highest J-Score, but with the constraint that no single phoneme confusion type (e.g., P $\rightarrow$ B) could appear more than three times. This crucial step ensures the final test is not overly weighted towards a single, dominant type of error and provides a broader, more robust diagnostic profile.

\section{Results}

\subsection{3.1. The SimPhon Speech Test-25 Test Set: Diagnostic Performance}
The culmination of the six-phase computational pipeline was the derivation of the SimPhon Speech Test-25 test set, a phonetically balanced list of 25 diagnostically powerful minimal word pairs. Each pair was selected based on its best-performing direction in the simulated diagnostic testing, yielding items optimized for high-frequency hearing loss.

Table 1 showcases the final SimPhon Speech Test-25 test set, meticulously ranked by its Youden's J statistic, a measure of diagnostic power derived from the simulated responses of both normal-hearing (NH) and hearing-impaired (HI) listeners. Notably, the set includes several pairs achieving perfect (J=1.0) or near-perfect (J=0.9) scores in simulation, specifically designed to target prevalent confusions like S$\rightarrow$F, T$\rightarrow$D, and P$\rightarrow$B. Furthermore, the diversity algorithm employed in Phase 6 was instrumental in ensuring the inclusion of other significant phonetic contrasts, such as CH$\rightarrow$T and D$\rightarrow$T, even if their individual J-Scores were slightly lower. This algorithmic constraint was vital for developing a comprehensive test instrument capable of probing a wide spectrum of phonetic challenges, thereby avoiding over-representation of a limited number of error types.

\begin{table}[h!]
\centering
\caption{The Final SimPhon Speech Test-25 Test Set, Ranked by Diagnostic Score (J)}
\begin{tabular}{|l|l|l|l|l|l|}
\hline
Stimulus & TargetConfusion & PhonemeChange & J\_Score & Sensitivity & Specificity \\
\hline
"sealing" & "feeling" & S $\rightarrow$ F & 1 & 1 & 1 \\
"ceilings" & "feelings" & S $\rightarrow$ F & 1 & 1 & 1 \\
"retlin" & "redlin" & T $\rightarrow$ D & 1 & 1 & 1 \\
"pepple" & "pebble" & P $\rightarrow$ B & 0.9 & 1 & 0.9 \\
"crap" & "crab" & P $\rightarrow$ B & 0.9 & 0.9 & 1 \\
"stet" & "stead" & T $\rightarrow$ D & 0.8 & 1 & 0.8 \\
"powter" & "powder" & T $\rightarrow$ D & 0.8 & 0.8 & 1 \\
"peeps" & "beeps" & P $\rightarrow$ B & 0.6 & 0.6 & 1 \\
"chop" & "top" & CH $\rightarrow$ T & 0.6 & 1 & 0.6 \\
"poppies" & "copies" & P $\rightarrow$ K & 0.4 & 0.9 & 0.5 \\
"dispersive" & "discursive" & P $\rightarrow$ K & 0.4 & 0.5 & 0.9 \\
"mund" & "munt" & D $\rightarrow$ T & 0.4 & 0.4 & 1 \\
"prone" & "crone" & P $\rightarrow$ K & 0.3 & 0.4 & 0.9 \\
"pus" & "puff" & S $\rightarrow$ F & 0.2 & 0.7 & 0.5 \\
"podge" & "pod" & JH $\rightarrow$ D & 0.2 & 1 & 0.2 \\
"cupid" & "cupit" & D $\rightarrow$ T & 0.1 & 0.1 & 1 \\
"mace" & "mase" & S $\rightarrow$ Z & 0.1 & 0.8 & 0.3 \\
"unheard" & "unhurt" & D $\rightarrow$ T & 0.1 & 0.1 & 1 \\
"brained" & "craned" & B $\rightarrow$ K & 0 & 0 & 1 \\
"kembel" & "pemble" & K $\rightarrow$ P & 0 & 0 & 1 \\
"brothers" & "crothers" & B $\rightarrow$ K & 0 & 0 & 1 \\
"laughter" & "laster" & F $\rightarrow$ S & 0 & 0 & 1 \\
"retail" & "resale" & T $\rightarrow$ S & 0 & 0 & 1 \\
"whittle" & "whistle" & T $\rightarrow$ S & 0 & 0 & 1 \\
"turning" & "churning" & T $\rightarrow$ CH & 0 & 0 & 1 \\
\hline
\end{tabular}
\end{table}

\subsection{Detailed Phoneme Error Analysis}
Following the word-level evaluation, a granular phoneme-level analysis was conducted to quantify the specific types of errors (deletions, insertions, and substitutions) made by the Automatic Speech Recognition (ASR) system when simulating both normal-hearing (NH) and hearing-impaired (HI) listeners. This approach allows for a deeper understanding of the perceptual mechanisms impacted by the simulated hearing loss.

\subsubsection{Overall Error Rates and Phoneme Operations}
The overall phoneme-level dissimilarity between the stimulus words and the ASR output was quantified using the Levenshtein distance. The average Levenshtein distance for the simulated Hearing-Impaired (HI) listener was 3.90, significantly higher than that for the Normal-Hearing (NH) listener, which was 3.08. This quantitative difference underscores the substantial impact of the simulated profound hearing loss and noise on phoneme recognition, confirming the ASR system's sensitivity to acoustic degradation.

A detailed breakdown of the most frequent phoneme-level operations revealed distinct patterns for each listener type.

\textbf{Phoneme Deletions:}
The most frequently deleted phonemes for both HI and NH listeners showed considerable overlap:
\begin{itemize}
    \item \textbf{HI Top 5 Deletions:} S, AH, IH, K, N
    \item \textbf{NH Top 5 Deletions:} AH, N, S, IH, K
\end{itemize}
The high prevalence of /S/ deletion in both conditions is consistent with its high-frequency spectral content, making it vulnerable to masking by noise and high-frequency hearing loss. Similarly, the frequent deletion of the vowels /AH/ and /IH/ and the nasal /N/ suggests that phonemes with inherently lower energy or shorter durations are consistently more susceptible to being missed under challenging listening conditions, irrespective of the hearing status. While the \textit{types} of phonemes frequently deleted are similar across both listener groups, the \textit{absolute counts} of these deletions are generally higher for the HI listener, reflecting a more severe loss of information.

\begin{figure}[h!]
    \centering
    \includegraphics[width=0.8\textwidth]{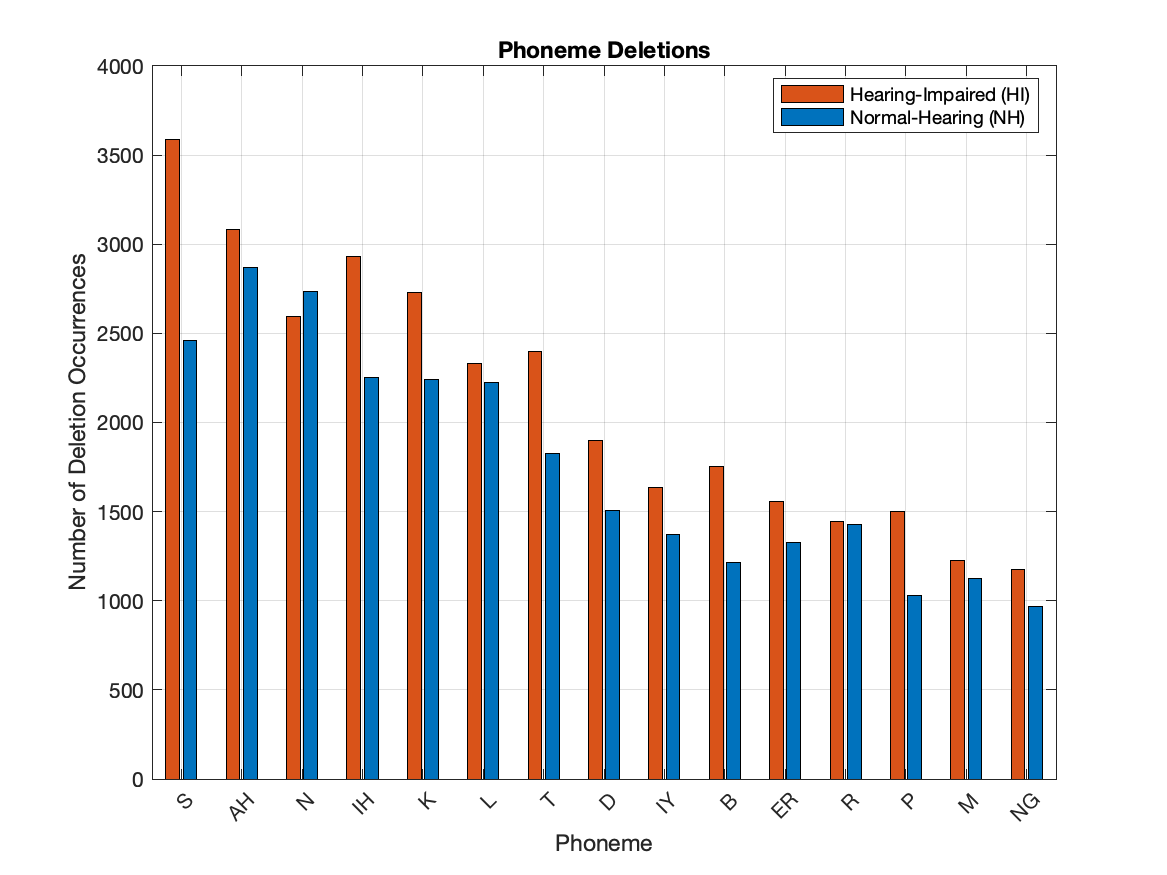}
    \caption{Comparison of the frequency of phoneme deletions for simulated Hearing-Impaired (HI) and Normal-Hearing (NH) listeners. Higher bars indicate more frequent deletion of a given phoneme.}
    \label{fig:phoneme_deletions}
\end{figure}

\textbf{Phoneme Insertions:}
Phoneme insertions, representing spurious phonemes added by the ASR system, occurred less frequently than deletions or substitutions. The top 5 insertions were:
\begin{itemize}
    \item \textbf{HI Top 5 Insertions:} IH, AH, N, HH, G
    \item \textbf{NH Top 5 Insertions:} T, B, IH, AH, G
\end{itemize}
The inserted phonemes are often short, relatively neutral vowels (/IH/, /AH/) or common consonants (/N/, /G/, /T/, /B/). This suggests that when the ASR system is highly uncertain due to degraded input, it may "hallucinate" simple, acoustically plausible phonemes. The presence of /HH/ (aspirated H) in the HI insertions could be indicative of the ASR identifying general breath or noise-like energy that it erroneously interprets as a phoneme in the absence of clear signal information. While insertions are fewer, they provide insight into the ASR's "fallback" recognition strategies under severe distortion.

\begin{figure}[h!]
    \centering
    \includegraphics[width=0.8\textwidth]{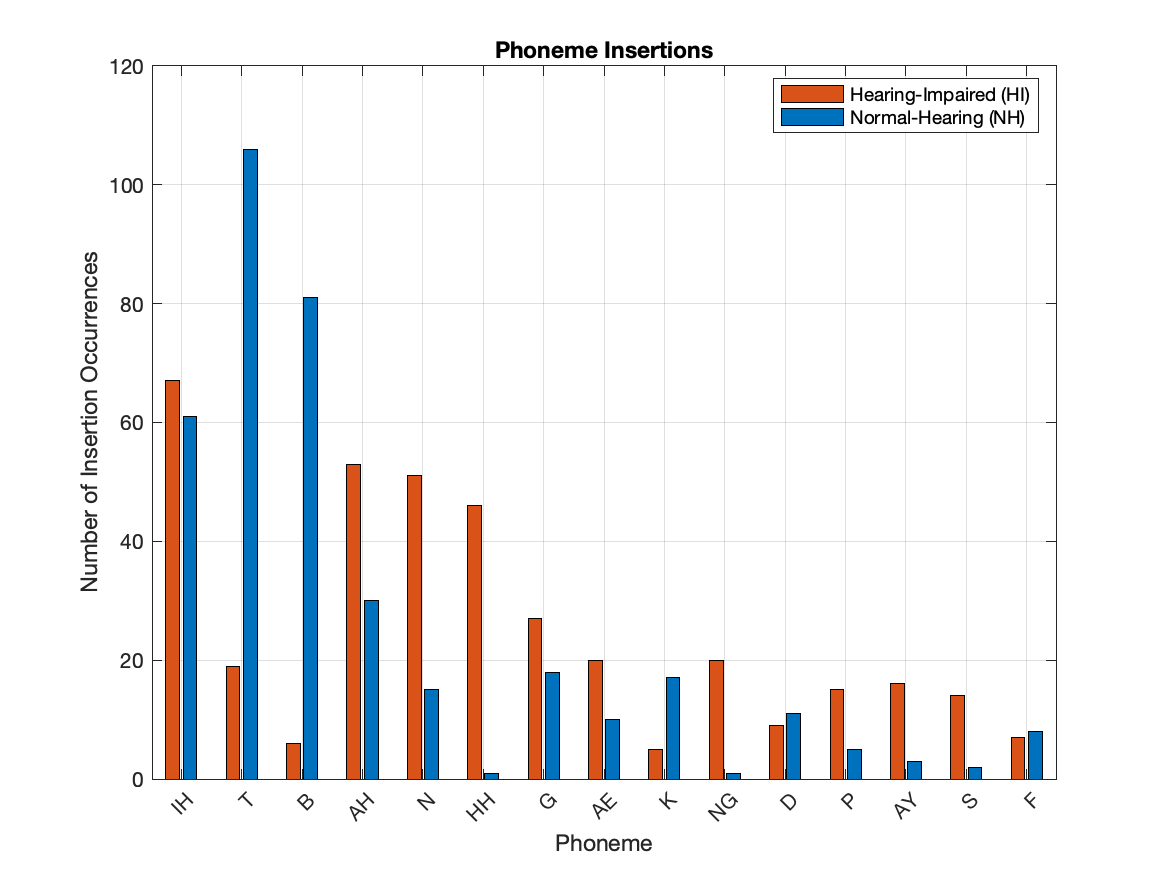}
    \caption{Comparison of the frequency of phoneme insertions for simulated Hearing-Impaired (HI) and Normal-Hearing (NH) listeners. Higher bars indicate more frequent insertion of a given phoneme by the ASR system.}
    \label{fig:phoneme_insertions}
\end{figure}

\textbf{Phoneme Substitutions:}
The patterns of phoneme substitution reveal the most diagnostically relevant differences between the simulated HI and NH listeners.
\begin{itemize}
    \item \textbf{HI Top 5 Substitutions:} T $\rightarrow$ OW, K $\rightarrow$ G, ER $\rightarrow$ AH, D $\rightarrow$ OW, IY $\rightarrow$ HH
    \item \textbf{NH Top 5 Substitutions:} S $\rightarrow$ F, T $\rightarrow$ D, D $\rightarrow$ T, CH $\rightarrow$ T, B $\rightarrow$ P
\end{itemize}
For the NH listener, the most common substitutions (e.g., S $\rightarrow$ F, T $\rightarrow$ D, B $\rightarrow$ P) primarily involve changes in voicing or manner of articulation, particularly for fricatives and plosives. These are well-documented types of confusions in human speech perception in noise, often associated with the loss of high-frequency spectral cues or temporal fine structure. The bidirectional confusion between /T/ and /D/ (and /B/ and /P/) is characteristic of voicing errors, while /S/ to /F/ is a common fricative confusion. These patterns for the NH listener align with expected phonological errors under moderate noise.

In stark contrast, the HI listener's top substitutions present a more complex and unexpected profile. Confusions such as T $\rightarrow$ OW, D $\rightarrow$ OW, ER $\rightarrow$ AH, and IY $\rightarrow$ HH suggest a severe breakdown in the acoustic distinctions between consonants and vowels/diphthongs, or between distinct vowel sounds. These types of errors are less typical of direct phonological substitutions seen in human perception and could indicate:
\begin{itemize}
    \item \textbf{Extreme Degradation:} The simulated profound hearing loss (85 dB attenuation at high frequencies) combined with challenging 5 dB SNR might lead to such severe spectral smearing or complete loss of crucial acoustic features that the ASR system's internal representations collapse, leading it to 'guess' basic vowel or diphthong-like sounds where consonants or other vowels were present.
    \item \textbf{ASR Model Behavior:} These patterns might also highlight specific sensitivities or limitations of the \texttt{wav2vec2.0} ASR model itself when confronted with highly aberrant input that falls outside its typical training distribution. For instance, the ASR might default to a vowel/diphthong when consonant cues are entirely absent, or it might be picking up residual low-frequency energy that it interprets as a vocalic element.
\end{itemize}
The visual representations in the phoneme confusion matrices (e.g., "HI Phoneme Confusion Matrix (Stimulus vs. ASR Output)" and "NH Phoneme Confusion Matrix (Stimulus vs. ASR Output)") graphically illustrate these divergent substitution patterns, clearly showing the density of incorrect recognition from a stimulus phoneme to a recognized phoneme. These matrices are crucial for a comprehensive understanding of the phoneme-level impact.

\begin{figure}[h!]
    \centering
    \includegraphics[width=0.8\textwidth]{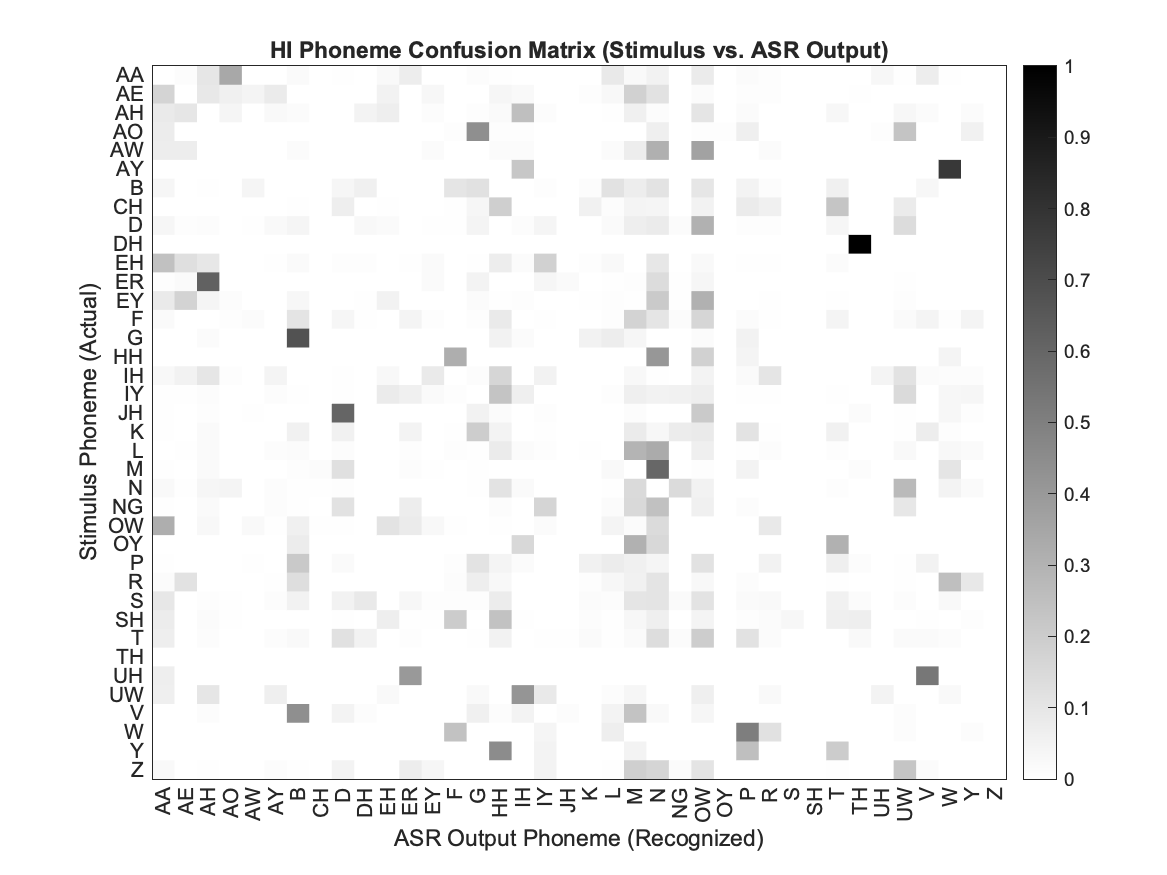}
    \caption{Phoneme confusion matrix for the simulated Hearing-Impaired (HI) listener, showing the proportion of times a stimulus phoneme (rows) was recognized as a particular ASR output phoneme (columns). Darker cells indicate higher probabilities of confusion.}
    \label{fig:hi_confusion_matrix}
\end{figure}

\begin{figure}[h!]
    \centering
    \includegraphics[width=0.8\textwidth]{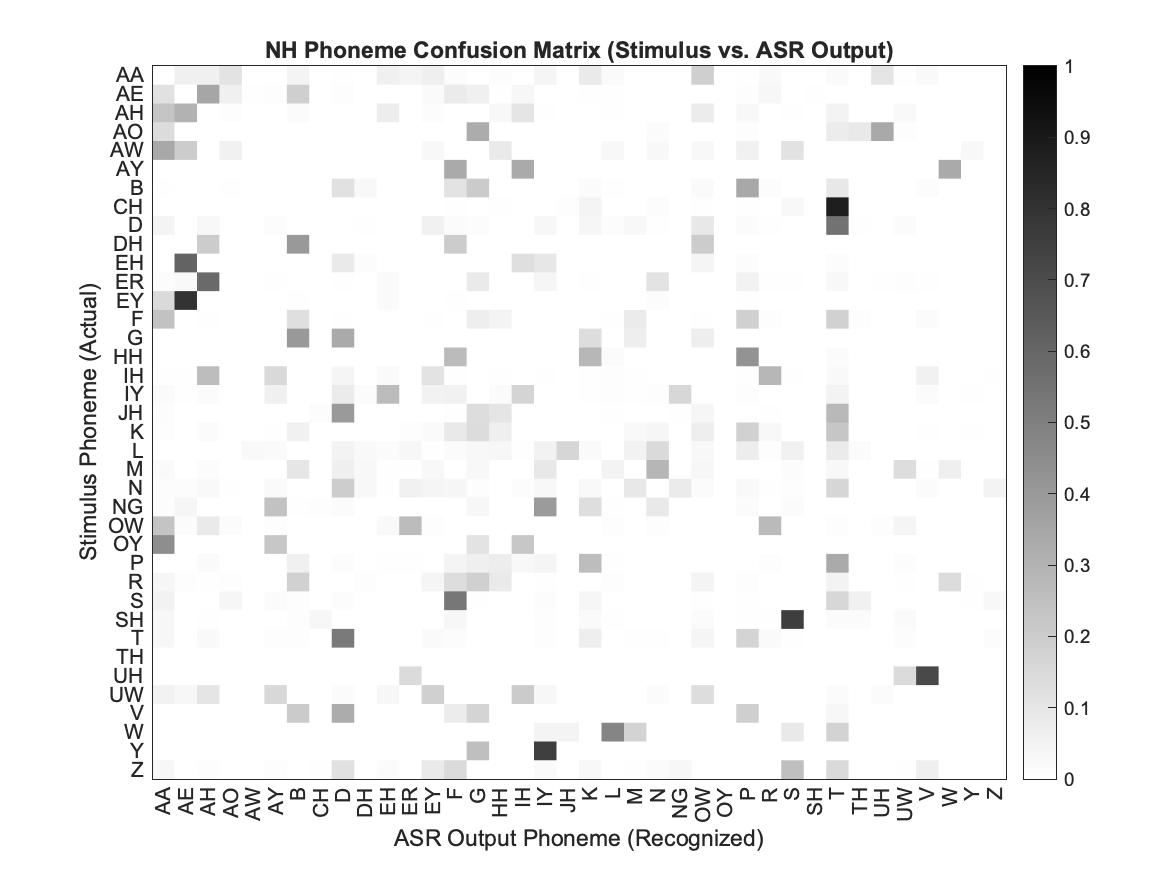}
    \caption{Phoneme confusion matrix for the simulated Normal-Hearing (NH) listener, showing the proportion of times a stimulus phoneme (rows) was recognized as a particular ASR output phoneme (columns). This matrix provides a baseline for comparison with HI errors.}
    \label{fig:nh_confusion_matrix}
\end{figure}

\section{Discussion}

\subsection{Interpretation of Results and Novel Contributions}
The SimPhon Speech Test methodology, leveraging an Automatic Speech Recognition (ASR) system as a proxy for human perception, successfully designed and validated a phonetically balanced minimal-pair speech test. The final SimPhon Speech Test-25 test set demonstrates strong diagnostic potential in simulated environments, effectively discriminating between normal and hearing-impaired conditions. This \textit{in silico} approach significantly de-risks and accelerates the development process, ensuring that only the most promising items are carried forward to more resource-intensive human trials. The SimPhon Speech Test-25 is not a random collection but a computationally optimized set tailored to elicit diagnostically relevant confusions.

The detailed phoneme error analysis further validates the SimPhon Speech Test methodology's capacity to reveal specific and quantifiable impacts of simulated hearing loss on speech recognition. The distinct error profiles observed between the HI and NH listeners, particularly in the substitution patterns, provide strong evidence that the SimPhon Speech Test captures more than just overall word intelligibility. The NH errors broadly reflect known human phonetic confusions in noise, while the HI errors point to more profound, ASR-specific distortions arising from severe signal degradation. This granular insight into phoneme-level confusions, deletions, and insertions is a novel contribution, offering a deeper understanding of how acoustic information is lost or distorted under conditions of hearing impairment.

\subsection{Comparison with Standard Audibility Models}
A key question addressed by this work was whether the SimPhon Speech Test provides diagnostic information beyond what could be predicted by standard, audibility-based models like the Speech Intelligibility Index (SII) (ANSI S3.5-1997, 1997). To test this, a correlation analysis was performed, comparing the SimPhon Speech Test's simulated diagnostic performance (Youden's J-Score) with the SII-predicted change in audibility (SII\_Normal - SII\_Impaired) for each SimPhon Speech Test-25 stimulus word under both normal and mild hearing loss profiles.

The analysis revealed no significant correlation between the SimPhon Speech Test's J-Score and the SII-predicted change in audibility (Pearson's R=-0.115, p=0.58 for a 'mild' simulation; R=0.097, p=0.6456 for a 'moderate' simulation). This is a critical finding. It suggests that the diagnostic power of the SimPhon Speech Test-25 pairs is not simply a function of word audibility. Instead, the SimPhon Speech Test's data-driven selection process, which is based on emergent phonetic confusions, has successfully identified items that are likely sensitive to the supra-threshold distortions and specific phonetic feature ambiguities that the SII framework does not fully capture. This provides strong evidence that the SimPhon Speech Test offers diagnostic insights that are complementary to standard audibility metrics.

\subsection{Insights from Acoustic Feature Analysis}
An acoustic analysis was performed to investigate whether the diagnostic efficacy of the word pairs could be explained by quantifiable global acoustic features of the stimuli. The analysis compared a group of "Good" diagnostically performing pairs (Youden's J>0.5) against a group of "Poor" pairs (J<0.1) from the final SimPhon Speech Test-25 test set. The absolute difference in five global acoustic features---spectral centroid, spectral skewness, spectral kurtosis, spectral flux, and harmonic ratio---is calculated for each word pair, and a two-sample t-test was used to compare the distributions of these differences between the two groups.

The results yielded no statistically significant difference for any of the tested features. For example, the difference in spectral centroid (p=0.92), spectral shape measures like skewness (p=0.58), or temporal change like spectral flux (p=0.49) showed no correlation with diagnostic power. This null result is a critical finding. It strongly suggests that the diagnostic potency of the SimPhon Speech Test-25 items is not determined by simple, global acoustic characteristics of the word stimuli.

This conclusion, taken together with the previously noted lack of correlation with the Speech Intelligibility Index (SII) and the distinct phoneme error patterns, reinforces the central hypothesis of this work. The data-driven, ASR-based selection process appears to have successfully isolated word pairs whose potential for confusion is driven by more subtle and complex cues than are captured by these broad acoustic metrics. The diagnostic power likely resides in fine-grained phonetic details, temporal dynamics, or formant transitions to which the ASR model is sensitive, aligning with the goal of developing a test that moves beyond simple audibility to probe specific perceptual deficits.

\subsection{Limitations and Future Work}
The primary limitation of this study is its reliance on ASR as a proxy for human listeners. While ASR models are increasingly sophisticated, their internal processing may not fully align with human auditory perception, particularly regarding complex supra-threshold distortions (Oxenham, 2008). Our simulation of hearing loss is purely acoustic (filtering and noise) and does not encompass the full physiological complexity of cochlear damage. However, this aligns with the audibility-centric view of mild-to-moderate hearing loss, where the loss of access to specific frequency cues is a dominant factor (Humes, 2007). A critical consideration is whether the observed HI-specific substitution patterns (e.g., consonants replaced by vowels) truly mirror human perception under profound hearing loss or are artifacts of the ASR's internal architecture and training data. While human listeners with severe hearing loss do experience significant speech distortion, a direct comparison with human phoneme confusion data under identical acoustic conditions would be necessary to fully validate these ASR-generated patterns. Nevertheless, the ASR's consistent and quantifiable errors provide a novel, data-driven means to explore the limits of speech perception in degraded conditions.

Several avenues for further analysis exist within the current data. A deeper dive into the phoneme-level deletions and insertions (especially beyond the top 5) could reveal more about what phonetic information is completely lost versus what is simply confused.

The immediate future work is the clinical validation of the SimPhon Speech Test-25 test set. A dedicated MATLAB script (SimPhon Speech Test\_Human\_Test.m) has been developed for this purpose, allowing for the collection of normative data from both normal-hearing and hearing-impaired listeners. This will allow us to compare the computationally-predicted performance with real-world human performance and to calibrate the test for clinical use. Investigating the integration of more advanced AI models, such as Speech Foundation Models (Zhou et al., 2025), for even more nuanced speech intelligibility prediction represents a promising future direction for refining this methodology. Specifically, understanding which acoustic features are lost, leading to the unique HI substitutions, can guide the design of stimuli that are particularly sensitive to these specific perceptual deficits. Furthermore, these patterns could be used to train or fine-tune ASR models to be more diagnostically aligned with human hearing loss. The visualizations (confusion matrices, deletion/insertion bar charts) serve as powerful tools for identifying these crucial patterns for clinical and research applications.

\section{Conclusion}
The SimPhon Speech Test methodology represents a successful proof-of-concept for the data-driven design of a frequency-specific speech test. By combining large-scale simulation with expert human curation and a principled selection algorithm, we have produced a phonetically balanced, 25-pair test computationally optimized for diagnosing the types of confusions common in high-frequency hearing loss. The lack of correlation with the SII model strongly suggests the test provides novel diagnostic information beyond simple audibility. This work lays the foundation for a new class of efficient and diagnostically powerful hearing assessment tools.


\begin{thebibliography}{99}
\bibitem{ANSI1997} ANSI S3.5-1997. (1997). \textit{Methods for Calculation of the Speech Intelligibility Index}. American National Standards Institute.

\bibitem{Bleeck2025} Bleeck, S. (2025). \textit{Advancing Hearing Assessment: An ASR-Based Frequency-Specific Speech Test for Diagnosing Presbycusis}. arXiv.

\bibitem{Hnath2016} Hnath, T. \& Van Handel, N. (2016). \textit{Phoneme and Word Scoring in Speech-in-Noise Audiometry}. ResearchGate.

\bibitem{Humes2007} Humes, L. E. (2007). The Importance of Speech Audibility. \textit{The ASHA Leader}.

\bibitem{Jean2025} Jean et al. (2025). Automating Speech Audiometry in Quiet and in Noise Using a Deep Neural Network. \textit{MDPI Biology}.

\bibitem{Kricos2006} Kricos, P. C. (2006). Audiologic management of older adults with hearing loss and dementia. \textit{American Journal of Audiology}.

\bibitem{Levenshtein1966} Levenshtein, V. I. (1966). Binary codes capable of correcting deletions, insertions, and reversals. \textit{Soviet Physics Doklady}.

\bibitem{Oxenham2008} Oxenham, A. J. (2008). Cochlear compression: implications for hearing aids and listeners with hearing impairment. \textit{Trends in Amplification}.

\bibitem{Plomp1978} Plomp, R. (1978). Auditory handicap of hearing impairment and the limited benefit of hearing aids. \textit{The Journal of the Acoustical Society of America}.

\bibitem{Wilson2005} Wilson, R. H. \& McArdle, R. (2005). Speech-in-noise tests: A look at the future. \textit{Seminars in Hearing}.

\bibitem{Zhou2025} Zhou, H., et al. (2025). Unveiling the Best Practices for Applying Speech Foundation Models to Speech Intelligibility Prediction for Hearing-Impaired People. \textit{arXiv}.

\end{thebibliography}
\end{document}